\documentclass{aa}  %% Astronomy & Astrophysics style class
\usepackage[varg]{txfonts}           %% A&A font choice

\usepackage{natbib}
\bibpunct{(}{)}{;}{a}{}{,} 
\usepackage{graphicx}
\usepackage[colorlinks=true,urlcolor=blue,citecolor=blue,linkcolor=blue]{hyperref}

\makeatletter
\newcommand{\bibnote}[2]{\@namedef{#1note}{#2}}
\newcommand{\biblink}[2]{\@namedef{#1link}{#2}}
\makeatother

\begin{document}

\title{First detection of AlF line emission towards M-type AGB stars}
%  \\
%  \thanks{Research supported in part by the US Air Force
   % under grant no. AFOSR-88-0285 and
    %the National Science Foundation under grant
    %no. DMS-85-21154}\fnmsep
 % \thanks{This is a second footnote}\\
%  resulting in asymptotically faster convergence\\
%  for the same amount of work per iteration}
%\subtitle{H$^{12}$CN and H$^{13}$CN excitation analysis in thecircumstellar outflow of R Scl\\}
% \author{M. Saberi\inst{1}
\author{M. Saberi  \inst{\ref{1},\ref{2}}
        \and T. Khouri \inst{\ref{3}}
       \and  L. Velilla-Prieto\inst{\ref{3}}
       \and  J. P. Fonfría\inst{\ref{4}}
        \and W. H. T. Vlemmings\inst{\ref{3}}
     \and  S. Wedemeyer\inst{\ref{1},\ref{2}}
      }
\institute{ Rosseland Centre for Solar Physics, University of Oslo, P.O. Box 1029 Blindern, NO-0315 Oslo, Norway  \label{1}
\\\email{maryam.saberi@astro.uio.no}
\and
Institute of Theoretical Astrophysics, University of Oslo, P.O. Box 1029 Blindern, NO-0315 Oslo, Norway \label{2}
\and
Dep. of Space, Earth and Environment, Chalmers University of Technology, Onsala Space Observatory, 43992 Onsala, Sweden
\label{3}
\and
Grupo de Astrofísica Molecular, Instituto de Física Fundamental, IFF-CSIC, C/ Serrano, 123, 28006, Madrid (Spain)
\label{4}
}
\date{}

\abstract
{The nucleosynthesis production of fluorine (F) is still a matter of debate. Asymptotic giant branch (AGB) stars are one of the main candidates for F production. However, their contribution to the total F budget is not fully known due to the lack of observations. 
In this paper, we report the detection of aluminium monofluoride (AlF) line emission, one of the two main carriers of F in the gas-phase in the outflow
of evolved stars, towards five nearby oxygen-rich (M-type) AGB stars.
We studied the Atacama large millimetre/sub-millimetre array (ALMA) observations of AlF ($v=0$, $J$\,=\,\mbox{4--3}, \mbox{9--8}, \mbox{10--9}, and \mbox{15--14}) and ($v=1$, $J$\,=\,\mbox{7--6}) line emission towards $o$ Ceti, and ($v=0$, $J$\,=\,\mbox{7--6} and \mbox{15--14}) lines towards R Leo. We also report a tentative detection of AlF ($v=0$, $J$\,=\,\mbox{7--6}) line in IK Tau, ($v=0$, $J$\,=\,\mbox{15--14}) line towards R Dor, and ($v=0$, $J$\,=\,\mbox{7--6} and $J$\,=\,\mbox{15--14}) lines in W Hya.
From spatially resolved observations, we estimated the AlF emitting region with a radius $\sim11R_{\star}$ for $o$ Ceti and $\sim9R_{\star}$ for R Leo.
From population diagram analysis, we report the AlF column densities of $\sim 5.8\times10^{15}$ cm$^{-2}$ and $\sim 3\times10^{15}$ cm$^{-2}$ for $o$ Ceti and R Leo, respectively, within these regions.
For $o$ Ceti, we used the C$^{18}$O ($v=0$, $J$\,=\,\mbox{3--2}) observations to estimate the H$_2$ column density of the emitting region. We found a fractional abundance of $f_{\rm AlF/H_2}\sim(2.5\pm1.7)\times10^{-8}$. This gives a lower limit on the F budget in $o$ Ceti and is compatible with the solar F budget $f_{\rm F/H_2}=(5\pm2)\times10^{-8}$.
For R Leo, a fractional abundance $f_{\rm AlF/H_2}=(1.2\pm0.5)\times10^{-8}$ is estimated.
For other sources, we cannot precisely determine the emitting region based on the available data. Assuming an emitting region with a radius of $\sim 11R_{\star}$ and the rotational temperatures derived for $o$ Ceti and R Leo, we crudely approximated the AlF column density to be $\sim(1.2-1.5)\times10^{15}$ cm$^{-2}$ in W Hya, $\sim(2.5-3.0)\times10^{14}$ cm$^{-2}$ in R Dor, and $\sim(0.6-1.0)\times10^{16}$ cm$^{-2}$ in IK Tau. These result in fractional abundances within a range of $f_{\rm AlF/H_2}\sim(0.1-4)\times10^{-8}$ in W Hya, R Dor, and IK Tau. }

\maketitle
%A&A Editorial Office: Astronomy & Astrophysics - Author?s guide 21

%%%%%%%%%%%%%%%%%%%%%%%%%%%%%%%%%%%%%%%%%%%%%%%%%%%%%%%%%%%%%
\section{Introduction}\label{Introduction}
%%%%%%%%%%%%%%%%%%%%%%%%%%%%%%%%%%%%%%%%%%%%%%%%%%%%%%%%%%%%%

Fluorine (F) is among the few elements whose cosmic origin is still the subject of debate.
It has only one stable isotope ($\rm ^{19}F$), which can be easily destroyed by proton, neutron, and alpha particle capture reactions in stellar interiors \citep[e.g.][]{Ziurys94,Abia15}. 
F is the most electronegative element, so it is extremely chemically reactive and can strongly bond to electron donors such as metals \citep{Ziurys94}.

There are several scenarios to explain the cosmic F production: (i) He-burning shell flashes in asymptotic giant branch (AGB) stars with initial masses of \mbox{$\sim$2--4} $M_{\odot,}$  subsequent thermal pulses, and third dredge-up; (ii) neutrino process occurring during supernova explosions; (iii) mergers between helium and carbon-oxygen white dwarfs; (iv) He-burning phase in Wolf-Rayet (WR) stars; (v) rapidly rotating massive stars \citep[e.g.][]{ Woosley95, Meynet00,Karakas10,Longland11, Abia15, Jonsson17, Limongi18, Ryde20}. A detailed overview of the F production sites and its role on the Galactic chemical evolution can be found in  \citet[][and references therein]{Grisoni20}.
The relative contributions of the aforementioned sites must be constrained by observations \citep[e.g.][]{Timmes95, Spitoni18, Olive19}.
Among all these candidates, AGB stars are the only sites of F production that have been observationally confirmed \citep{Jorissen92, Federman05, Werner05, Abia15, Abia19}.

The AGB phase is a late phase of evolution for stars with an initial mass of \mbox{1--8} $M_{\odot}$.
AGB stars play a significant role in the Galactic chemical evolution by ejecting of newly synthesised elements to the interstellar medium through strong stellar winds. 

Models of \cite{Lugaro04} and \cite{Karakas10} predict that the F enrichment is a strong function of the initial stellar mass and metallicity. For solar metallicity models, they predict the highest F formation in AGB stars with initial masses to be between \mbox{2 and 4} $M_{\odot}$ and for them to be maximal in the \mbox{3--3.5} $M_{\odot}$ mass range.

%Observationally, it has been shown that AGB stars do contribute to fluorine production \citep{Jorissen92, Abia15, Abia19}.
\cite{Jorissen92} reported a fluorine over-abundance up to 30 times the solar value in AGB stars using the infrared vibration-rotation lines of hydrogen fluoride (HF).
They also found that the fluorine enrichment is correlated with the enrichment of atomic carbon.
An over-abundance ranging from 10-250 times the solar abundance are reported in a number of hot post-AGB stars from far-UV observations of F V and F IV \citep{Werner05}. This significant over-abundance of F is believed to be the F synthesised during the preceding AGB phase, which is brought to the surface of the post-AGB star.

Previous determinations of F abundance in AGB stars are mainly based on near infrared observations of HF lines. However, significant contamination of these spectral lines with telluric lines in this wavelength region prevents accurate determination of the F abundance \citep[e.g.][]{Abia09, Abia10, Abia15}.

Chemical equilibrium models of F-bearing species predict a considerable amount of F to be locked into aluminium monofluoride (AlF) and HF in the outflow of AGB stars. %\citep{Agundez20}. 
%Therefore, the abundance of AlF and HF can provide a good estimate of the total F budget in the outflow of AGB stars.
%From chemical equilibrium models, 
The metal-containing molecules such as AlOF, CaF, and CaF$_2$ are also expected to be abundant at radii larger than $10R_{\star}$ in M-type AGB stars \citep{Agundez20}. However, these species have not been detected in the outflow of AGB stars yet.

The lines of rotational transitions of AlF and HF have been previously reported in the outflow around the carbon-rich AGB star IRC+10216 \citep{Ziurys94, Agundez11,Agundez12}.
They estimated relative abundances of $f_{\rm AlF/H_2}\sim1\times10^{-8}$ and $f_{\rm HF/H_2}\sim8\times10^{-9,}$ which results in a lower limit of $f_{\rm F/H_2}\sim2\times10^{-8}$ for the total F budget in IRC+10216.
This is consistent with the solar value of $f_{\rm AlF/H_2}=(5\pm2)\times10^{-8}$ reported by \cite{Asplund21}.
Moreover, \cite{Danilovich21} recently reported the detection of rotational lines of AlF ($v=0, J=7-6$) towards the S-type AGB star, W Aql. They estimated a fractional abundance of $f_{\rm AlF/H_2}=7.2\times10^{-8} - 1.0\times 10^{-7}$ in the inner CSE within a radius of $\sim15R_\star$ and $f_{\rm AlF/H_2}= 4.0\times10^{-8}$ within $\sim90R_\star$ using radiative transfer modelling. 
%Moreover, \cite{Danilovich21} recently reported the detection of rotational lines of AlF ($v=0, J=7-6$ and $v=1, J=7-6$) towards the S-type AGB star, W Aql. They estimated a fractional abundance of $f_{\rm AlF/H_2}=7.2\times10^{-8} - 1.0\times 10^{-7}$ in the inner CSE within a radius of $\sim15R_\star$ and $f_{\rm AlF/H_2}= 4.0\times10^{-8}$ within $\sim90R_\star$ using radiative transfer modelling to fit the $v=0$ line. 

Observations of multiple transitions of F-bearing species at sub-millimetre wavelengths can provide a more accurate determination of the total F budget in AGB stars.
In this paper, we report the first detection of rotational lines of AlF towards M-type AGB stars, $o$ Ceti and R Leo, with the Atacama large millimetre/sub-millimetre array (ALMA).
We also report tentative detection of AlF lines towards W Hya, R Dor, and IK Tau. % which are presented in App. \ref{App-2}. 

%===========================================================================
\begin{table}[]
\caption{M-type AGB stars with detected AlF line emission.}
  \centering
  \setlength{\tabcolsep}{1.0pt}
\begin{tabular}{cccccccccc}
\hline
Star &  $\dot M$ & d & $V_{\rm LSR}$ & $V_{\rm exp}$ & $R_{\star}$ &  $T_{\star}$  & P & $M_\star$\\
     & \tiny($10^{-7} M_{\odot} yr^{-1}$) & \tiny(pc) & \tiny(km s$^{-1}$)& \tiny(km s$^{-1}$) & \tiny(mas) & \tiny(K)  & \tiny(days) &\tiny($M_{\odot}$)
     \\ 
\hline
$o$ Ceti &  $1$ & 102 & 47 & 3 & 15 &  2800 & 332 & 1.0 \\
R Leo &  1 & 130 & 0.5 & 6 & 13.5 &2800  & 310 & 1.5 \\
W Hya &  1 & 104 & 41 & 7 & 20 &  2950 & 388 & 1.0 \\
R Dor & 1 & 45 & 6.9 & 6 & 27.5 &  2400 & 175$^\dag$ & 1.0-1.3 \\ %362/175
IK Tau & 80 & 265 & 34 & 18.5 & 6 & 2100  & 470 & 1.1-1.5\\
\hline
\end{tabular}
\label{Source}
\tablefoot{Columns 2 to 8, for the first four sources, are taken from \citet[][]{Vlemmings19} and the stellar masses for W Hya, R Dor, and IK Tau are from \cite{Danilovich17}. For $o$ Ceti and R Leo, they are estimated based on the $^{17}$O/$^{18}$O ratio presented by \cite{DeNutte17} and the O isotopic ratios are taken from \cite{Hinkle16}. For IK Tau, all parameters are taken from \citet[][and references therein]{Velilla17}. We note that $R_\star$ listed here are the measured radii in the near-infrared. \dag Period of R Dor varies between 175 and 362 days.} 
\end{table}
%===========================================================================

%%%%%%%%%%%%%%%%%%%%%%%%%%%%%%%%%%%%%%%%%%%%%%%%%%
\section{Observations}\label{Observations}                              %%
%%%%%%%%%%%%%%%%%%%%%%%%%%%%%%%%%%%%%%%%%%%%%%%%%%

We used observations of the ALMA main arrays in Bands 4, 6, 7, and 8 from projects 2016.1.00004.S, 2017.1.00191.S, 2018.1.00749.S, and 2018.1.00649.S for $o$ Ceti, 
projects 2016.1.01202.S and 2017.1.00862.S for R Leo, and project 2016.1.00119.S for IK Tau. 
In addition, we used ALMA compact array (ACA) observations in Band 8 from project 2018.1.01440.S for $o$ Ceti, R Leo, R Dor, and W Hya and project 2016.2.00025.S for W Hya.

%The ACA spectra are extracted from the central pixel.
Table \ref{Source} lists the observed sources and their physical  properties.
The observed lines, their spectroscopic parameters, date of observations, and measured intensities are listed in Table \ref{AlF}. We also investigated any possible blends from the Cologne database for molecular spectroscopy (CDMS) \citep{Muller01, Muller05} and jet propulsion laboratory (JPL) spectroscopic databases \citep{Pickett98} (see notes in Table \ref{AlF}).

%We note that the AlF $v=0$, $J$\,=\,\mbox{15--14} line is also covered in the Band 8 ACA observations of stars TX Psc, TW Hor, EP Aqr, U Hya, and R Hya, but it was not detected towards any of them.

%We should note that our Band 8 ACA observations include five more sources Tx Psc, TW Hor, EP Aqr, U Hya, and R Hya in which cover AlF ($v=0, J=15-14$) at 494.2 GHz but we have no detection for these sources.

The data calibration was performed following the standard ALMA procedures with the Common Astronomy Software Application (CASA) \citep{McMullin07}. % and GILDAS\footnote{https://www.iram.fr/IRAMFR/GILDAS} software.
Details of the data processing can be found in \cite{Khouri18, Vlemmings19, Fonfria19}.
The calibration uncertainty will depend on the flux calibrator used and typically ranges from 5-20$\%$ \citep{Francis20}. In this paper, we assumed a typical $10\%$ uncertainty for the flux calibration. The integrated flux uncertainties that are listed in column 7 in Table \ref{AlF} include the uncertainties from flux calibration and from Gaussian fitting on the lines.

%In our observations, the four spectral windows cover frequency range 479.09-480.97, 480.97-482.84, 491.15-493.03, 493.04-494.92 GHz.

%===========================================================================
\begin{table*}[t]
  \centering
  \setlength{\tabcolsep}{2.2pt}
    \caption{Detected AlF emission lines towards M-type AGB stars by ALMA observations.}
  \begin{tabular}{@{} ccllcccccccccc@{}}
\hline
 Source & $\nu$  & Transition & $E_{\rm u}$ & $g_u$  & $A_{ul}$  & Flux & FWHM  & $\rm V_c$ & Ang. res. &  M.R.S. & Apr. & Obs. Date & $\phi$\\
   & & \tiny (GHz)  &  & \tiny (K) & & \tiny ($10^{-3}  s^{-1}$)  & \tiny (Jy km s$^{-1}$) & \tiny (km s$^{-1}$) &\tiny  (km s$^{-1}$)  &(arcsec)  & (arcsec) & (arcsec) &\\
\hline
\hline
$o$ Ceti  & 131.8988 &$v=0, J=4-3$& 16 & 9 & 0.025 & $0.12\pm0.04$ & 5.9 &  47.8 & 0.077 & 2.897 & 0.25 & 21-09-17 & 0.9 \\
          & 228.7165 &$v=1, J=7-6$& 1184 & 15 & 0.147 & $0.46\pm0.14$ & 9.7 & 48.5 & 0.049 & 1.267 & 0.5 & 22-09-17 & 0.9 \\
          & 296.6988 &$v=0, J=9-8$& 71 & 19 & 0.299 & $1.84\pm0.23$ & 6.6 & 47.7 & 0.286 & 9.238 & 0.5 & 19-11-18 & 0.1\\
          & 329.6416$^*$ &$v=0, J=10-9$& 87 & 21 & 0.412 & $4.39\pm0.59$ & 8.7 & 47.2 & 0.022 & 0.570 & 0.5 & 09-11-17 & 0.8 \\
          & 494.2268 &$v=0, J=15-14$& 190 & 31 & 1.41 & $7.60\pm0.84$ & 6.2 & 47.4 &  0.172 & 11.962 & 0.5 & 28-11-18 & 0.1\\
\hline
R Leo & 230.7938 & $v=0, J=7-6$ & 44 & 15 & 0.138 & $0.24\pm0.06$ & 7.2 & -1.0 & 0.133 & 1.941 & 1 & 01-10-16 & 0.5\\
&        &              &      &    &       &  $0.26\pm0.04$        & 6.4 & 0.3  & 1.275 & 13.166 & 5 & 22-03-17 & 0.6\\
&        &              &      &    &       &  $0.22\pm0.03$        & 5.8 & 0.2   & 0.365 & 5.521 & 2 & 03-05-17 & 0.0\\
&        &              &      &    &       &  $0.15\pm0.04$        & 6.2 & -2.3  & 0.028  & 0.695 & 0.3 & 21-09-17 & 0.7\\
&        &              &      &    &       &  $0.24\pm0.05$     & 4.3 & -0.6 & 0.022  & 0.431 & 0.3 & 03-10-17 & 0.7\\
&        &              &      &    &       &  $0.19\pm0.05$ & 5.7 & -0.6   & 0.025  & 0.597 & 0.3 & 27-10-17 & 0.9 \\
      & 494.2268 &$v=0, J=15-14$& 190 & 31 & 1.41 &  $3.16\pm0.49$ & 5.7 & 0 & 2.570 & 15.118 & CP & 24-03-19 & 0.2\\
\hline
W Hya & 230.7938$^{**}$ & $v=0, J=7-6$ & 44 & 15 & 0.138 & $0.49\pm0.15$ & 11 & 39.0 & 5.089 & 29.537& CP & 21-05-17 & 0.5\\
& 494.2268 &$v=0, J=15-14$& 190 & 31 & 1.41 & $4.29\pm1.22$ & 7.3 & 41.1 & 2.039 & 13.634 & CP & 08-06-19 & 0.8 \\
\hline
R Dor & 494.2268 &$v=0, J=15-14$& 190 & 31 & 1.41 & $2.13\pm0.72$ & 7.7 & 5.5 & 2.135 & 13.812 & CP & 08-06-19 & 0.3\\
\hline
IK Tau & 230.7938$^{**}$ & $v=0, J=7-6$ & 44 & 15 & 0.138 &  $0.23\pm0.04$ & 13.7 & 36.0 & 0.752 & 8.412 & 3 & 13-04-17 & -\\

\hline
 \end{tabular}
\tablefoot{Spectroscopic data are taken from the CDMS. Integrated flux, the full width at half maximum (FWHM), and the line central velocity (V$_{\rm c}$) are from Gaussian fitting. Flux uncertainties ($\sigma_{Flux}$) are a summation of Gaussian fitting uncertainty and $10\%$ of the total flux due to calibration uncertainty of ALMA data. MRS stands for maximum recoverable scale. Apr. denotes circular aperture used to extract the spectra, and CP denotes where the spectra are extracted from the central pixel ($6.6^{\prime\prime} \times 4.4^{\prime\prime}$) for the ACA observations. $\phi$ denotes the stellar variability phase at visual wavelengths where $\phi_{min}=1$ and $\phi_{max}=0$. ${}^*$ and ${}^{**}$ lines are likely blended with an SO$_2$ line at 329.6459 GHz and a ${}^{50}$TiO$_2$ line at 230.7931 GHz, respectively.}
 \label{AlF}
\end{table*}
%===========================================================================
%1. 

%===========================================================================
\section{Excitation analysis and results}\label{Analysis}
%===========================================================================
%%===========================================================================
\begin{figure*}[tb]
 \centering
\includegraphics[width=180mm]{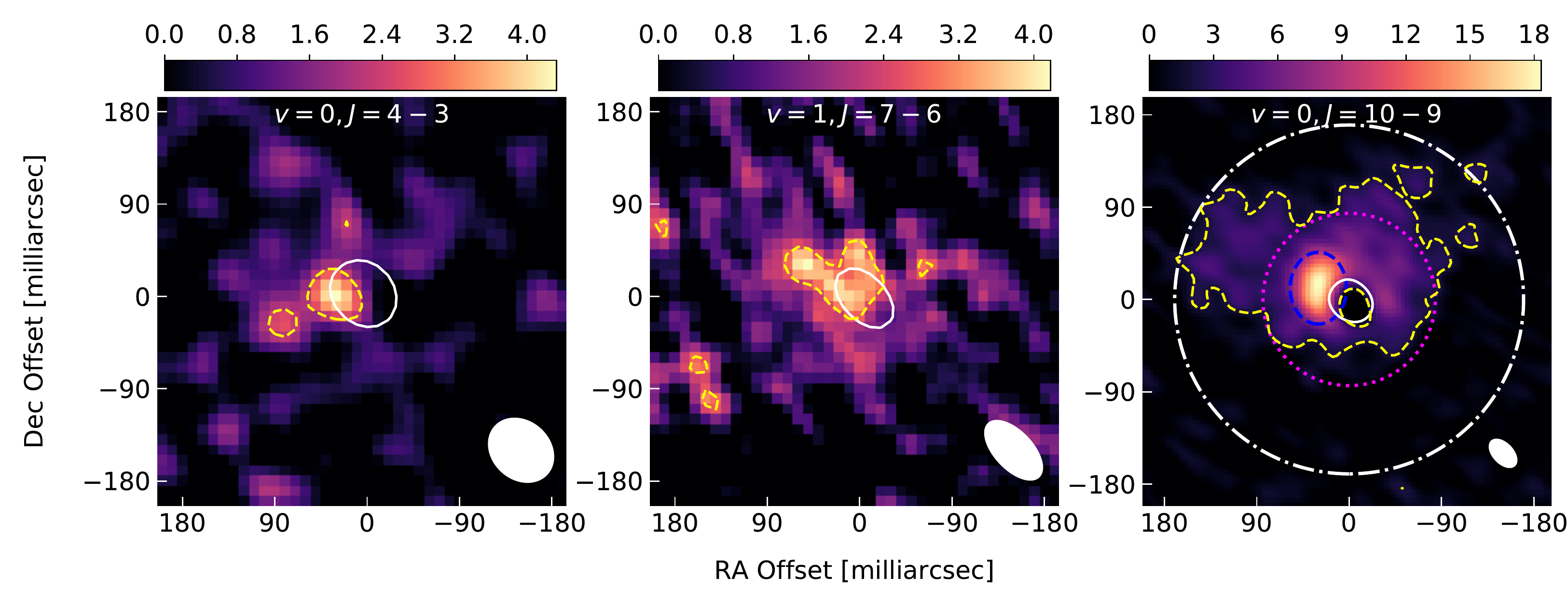} 
 \caption[]{Integrated emission of the spatially resolved lines of AlF towards $o$~Ceti. The lines are integrated in the 131.873-131.883 GHz (41-54 km s$^{-1}$) range for the $v$\,=0\,, $J$\,=\,\mbox{4--3} line, 228.675-228.685 GHz (36-59 km s$^{-1}$) for $v$\,=1\,, $J$\,=\,\mbox{7--6}, and 329.606-329.577 GHz (32-59 km s$^{-1}$) for the $v$\,=0\,, $J$\,=\,\mbox{10--9} line. The scale is given in Jy~km/s/beam. The lines shown are indicated at the top of each panel. The white contours show the 50\% stellar continuum emission level at the corresponding frequencies and the dashed yellow contours mark the 3-$\sigma$ level of the line emission. The filled white ellipses indicate the beam size in each observation. In the right panel, the regions over which the lines were integrated (see text) are indicated by the dashed blue ellipse (peak AlF emission), dotted magenta circle (CO~$v=1,J=3-2$ line emission region), and dotted-dashed white circle (all AlF emission lines).}
\label{Mira-AlF-inner}  
\end{figure*}
%%=========================================================================

%===========================================================================
\subsection{$o$ Ceti}\label{Mira}
%===========================================================================

%We detect four rotational lines within the ground vibrational state $v=0$, which are $J$\,=\,\mbox{4--3}, \mbox{9--8}, \mbox{10--9}, and \mbox{15--14}, and the $J$\,=\,\mbox{7--6} line in the vibrationally excited state $v=1$ (Table~\ref{AlF}). 
\subsubsection{AlF population diagram}\label{AlFinMira}

We identify four rotational lines of AlF within the ground vibrational state $v$\,=\,0, $J$\,=\,\mbox{4--3}, \mbox{9--8}, \mbox{10--9}, and \mbox{15--14}, and the vibrationally excited line $v$\,=\,1, $J$\,=\,\mbox{7--6,} which are listed in Table~\ref{AlF}. Rotational transition frequencies of AlF are taken from CDMS \citep{Wyse70,Hoeft70}.
Figure \ref{Mira-AlF-inner} presents the integrated flux density of spatially resolved AlF lines towards $o$ Ceti. 
From the $J$\,=\,\mbox{10--9} line, the emitting region is estimated to have a diameter of $\sim0.34^{\prime\prime}\sim 34$ au $\sim 22 R_{\star}$ based on the $3\sigma$ emission region, where $R_\star$ refers to the stellar radius measured in infrared.
%(see Fig.~\ref{Mira-AlF-inner})

We derived the column density of AlF towards these three regions:
1: the region with a diameter of $\sim0.34^{\prime\prime} (\sim 22 R_\star)$ centred on the star shown in Fig.~\ref{Mira-AlF-inner}, rightmost panel, shown by a dotted-dashed white circle, encompassing all five observed AlF lines; 
2: a circular region centred on the star with a diameter of $0.168^{\prime\prime} (\sim 11R_\star)$ shown in Fig.~\ref{Mira-AlF-inner}, rightmost panel, represented by a dotted magenta circle, where we use the three spatially resolved lines ($v$\,=0\,, $J$\,=\,\mbox{4--3,}  \mbox{10--9}, and $v$\,=1\,, $J$\,=\,\mbox{7--6});
3: a small elliptical region at north-east of the star with a diameter of $0.055^{\prime\prime}\times0.07^{\prime\prime} (\sim 4 R_\star)$ shown in Fig.~\ref{Mira-AlF-inner}, rightmost panel, represented by a dashed blue ellipse, where we could only use the AlF ($v$\,=0\,, $J$\,=\,\mbox{10--9,} and $v$\,=1\,, $J$\,=\,\mbox{7--6)} lines. The $v$\,=0\,, $J$\,=\,\mbox{4--3} extracted from this region was too weak to be used for the analysis in region 3.

For regions 1 and 2, we used the population diagram method \citep{Goldsmith99} to estimate the AlF rotational temperature ($T_{\rm rot}$) and mean column density ($N_{\rm AlF}$). 
This is a reasonable approximation since AlF lines are optically thin, as we show later, and AlF arises from a relatively small region ($<11R_\star$) in the inner part.
Our analysis also shows that the rotational temperature does not vary significantly between regions 1 and 2. 
The rotational temperature describes the excitation of the rotational levels of the molecule, and in the case of local thermodynamic equilibrium (LTE), it also represents the kinetic temperature of the gas.
For region 3, we were only able to use the $J$\,=\,\mbox{10--9} line in the ground vibrational state, and a population diagram analysis was not possible.

The population level of a given molecule follows

\begin{equation}
ln(\frac{N_u}{g_u}) = ln(\frac{N_{\rm AlF}}{Z})- \frac{E_u}{k_{\rm B} T_{exc}},
\label{NvJ}
\end{equation}
where $N_u$ and $g_u$ are the column density and statistical weight of the upper level, respectively, $N_{\rm AlF}$ is the total column density of the AlF molecules, $Z$ is the partition function, $E_{\rm u}$ is the energy of the upper level, $k_{\rm B}$ is the Boltzmann constant, and
$T_{\rm exc}$ is the excitation temperature.
%$Z(T_{\rm rot})$ is the rotational partition function, $E_\rm u$ is the energy of the upper level. $Z(T_{\rm rot})$ is calculated using: 
%\begin{equation}
%Z_{\rm rot}(T_{\rm rot}) = \sum_{J}(2J+1)e^{-J(J+1)B/k_B T_{\rm rot}},
%\end{equation}
%where we considered that $J$ runs from 0 to 94 and B=16488.355 MHz, which was taken from CDMS.

We assumed optically thin emission in the calculation of $N_{\rm u}$,
which is estimated using
\begin{equation}
N_u = \left(\frac{4\pi d^2}{\pi r_e^2}\right)\left(\frac{W}{A_{ul}h\nu}\right),
\label{NvJ}
\end{equation}
where the first term accounts for the geometrical dilution in which $d$ is the distance to the source and $r_e$ is the radius of the emitting region. $W$ is the flux of the line in units (W m$^{-2}$), and for each line it is individually extracted from the regions that are listed above. $A_{\rm ul}$ represents the Einstein-A coefficients, which express the probability of the spontaneous emission from the upper level $u$ to the lower level $l$.

\begin{figure}[t]
 \centering
\includegraphics[width=\columnwidth]{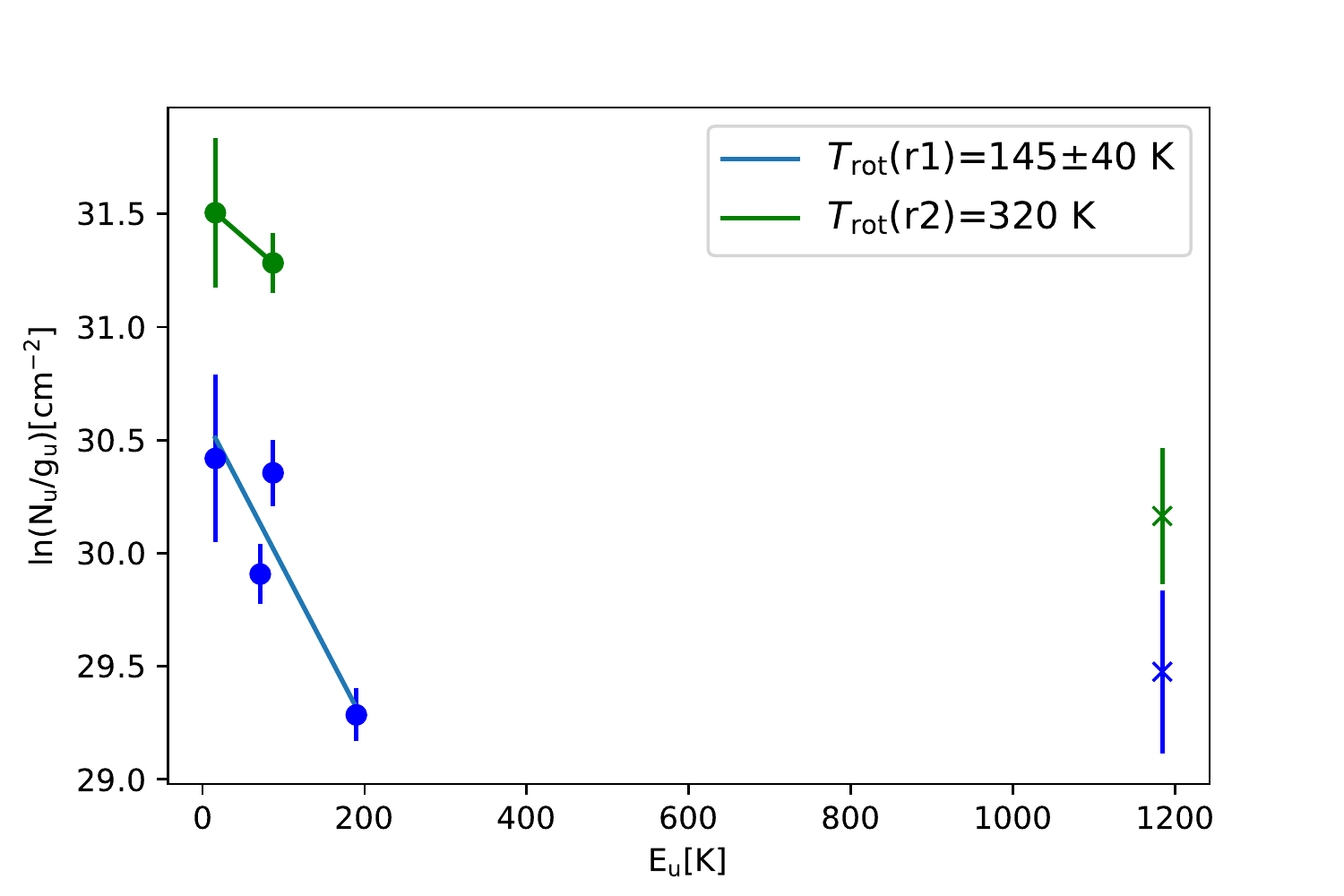} 
 \caption[]{\label{}
    Population diagram of observed AlF rotational transitions towards $o$ Ceti that are listed in Table \ref{AlF}. Blue points correspond to flux densities that are extracted from region 1 and green points correspond to those from region 2 (see Sect. \ref {Mira} for explanations).
    The vibrationally excited $v=1$ lines in both regions are also shown by x symbols, but they were not considered in the calculations of the rotational temperatures. The vertical error bars are calculated from the uncertainties of the integrated fluxes (uncertainty from the Gaussian fitting plus 10$\%$ of the total flux from the calibration of the data) and are considered in the fitting process.}
\label{Mira-RD}  
\end{figure}

 From the population diagram shown in Fig. \ref{Mira-RD}, we find $T_{\rm rot}=145\pm40$ K and $T_{\rm rot} = 320$ K for regions~1 and 2, respectively. The accuracy of the latter results is limited as we only have two points in the population diagram that correspond to rotational lines in $v$\,=0\,. 
Using $Z(145~{\rm K})=183.57$ and $Z(320~{\rm K})=404.73$, we find mean column densities $N_{\rm AlF}=(3.6\pm1.1)\times10^{15}$~cm$^{-2}$ and $N_{\rm AlF}=2.1\times10^{16}$~cm$^{-2}$ for regions 1 and 2, respectively, as listed in Table \ref{Mira-results} (Model A).

\begin{figure*}[tb]
 \centering
\includegraphics[width=170mm]{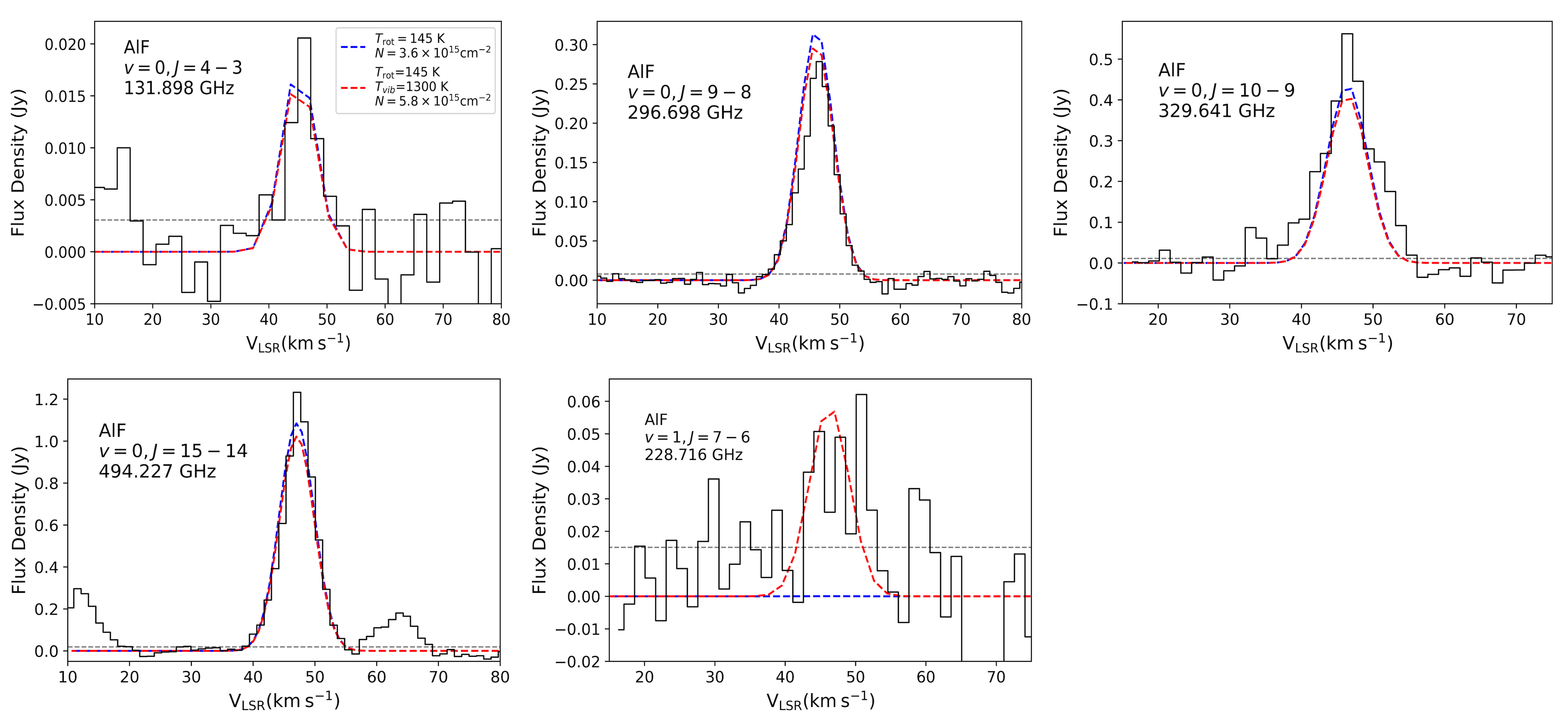}
 \caption[]{\label{}
    AlF lines observed by ALMA towards $o$ Ceti that are listed in Table \ref{Observations}  extracted from region 1 (black solid lines) overlaid with model results (dashed lines). Models indicate that two separate rotational and vibrational temperatures are need to reproduce all observed lines, which is discussed in Sect. \ref{Mira}. The line rest frequencies and transitions are marked in each panel. The grey dashed lines show the rms level.}
\label{Mira-AlF-R1}  
\end{figure*}

\begin{figure*}[tb]
 \centering
\includegraphics[width=170mm]{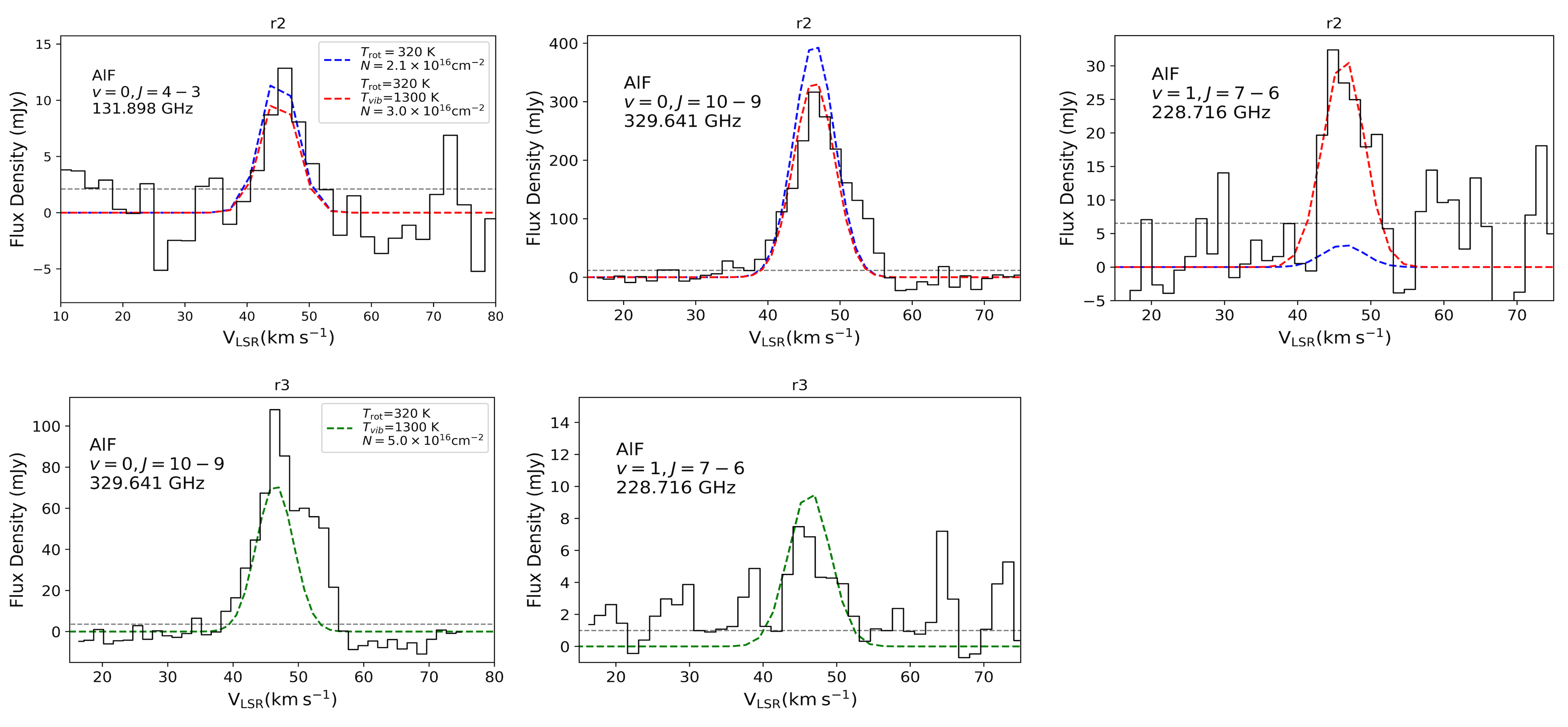} 
 \caption[]{\label{}
    AlF lines observed by ALMA towards $o$ Ceti extracted from region 2 with a diameter of $0.168^{\prime\prime} \sim 11 R_{\star}$ (upper panel) and region 3 with diameter of $0.055^{\prime\prime}\times0.07^{\prime\prime} \sim 4 R_{\star}$ (lower panel) shown with black solid lines overlaid with model results (dashed lines). Models indicate that two separate rotational and vibrational temperatures are needed to reproduce all observed lines, which is discussed in Sect. \ref{Mira}. The line rest frequencies and transitions are marked in each panel. The grey dashed lines show the rms level. We note that the asymmetric feature in the right wing of the $J=10-9$ line in region 3 might be due to a blend with a high excitation SO$_2$ line lying at 329.6459 GHz that becomes more prominent in the innermost regions of the circumstellar envelope.}
\label{Mira-AlF-R2-3}  
\end{figure*}

To test our assumption of optically thin emission, we estimated the optical depth at the line centre of all lines from all selected regions using
\begin{equation}
\tau_0=\frac{c^2 A_{ul} N_u}{8\pi \nu^2 \Delta\nu \sqrt{\pi}/2\sqrt{ln2}} \left(\exp\left(\frac{h\nu}{k_{\rm B} T_{\rm exc}}\right)-1\right).
\label{opacity}
\end{equation}
We find optical depths in the $0.09-0.5$ range in Region~1 and $0.1-0.5$ in Region~2, confirming our assumptions of emission that is not optically thick.
To investigate the accuracy of the excitation temperature and column density from the population diagram, we calculated the flux density ($S_\nu$) of the lines using
\begin{equation}
 S_v = N_{\rm AlF}(\pi r_e^2) \frac{g_u \exp(\frac{-E_u}{k_B T_{\rm rot}})} {Z_{\rm rot}(T_{\rm rot})} \frac{ A_{\rm ul} h \nu}{4 \pi d^2} \:\: \phi_{\nu},
\label{Sv}
\end{equation}
where $\phi_{\nu}=(exp(-(\nu-\nu_0)^2/\Delta\nu^2)/(\Delta\nu \sqrt\pi)$ is the line profile where $\Delta\nu=\nu_0 \Delta v/c,$ and we assumed $\Delta v=4$ km s$^{-1,}$ which fits the width of the observed AlF lines.
Using the derived column densities and rotational temperatures from the population diagrams in Eq. \ref{Sv} to reproduce the synthetic spectra in both regions, we significantly under-predicted the line strength of the vibrationally excited $v$\,=1\, line (see dashed blue lines in Fig. \ref{Mira-AlF-R1} for region 1 and upper panel of Fig. \ref{Mira-AlF-R2-3} for region 2). As can be seen, the derived temperature and column density only characterise the populations of the rotational levels, which also indicates the quasi-thermal excitation discussed above.
 This can be caused by different excitation mechanisms dominating the excitation of the $v$\,=0\, and $v$\,=1\, levels. The ground state excitation is most likely dominated by molecular collision, while the vibartionally excited state is mostly populated by the radiation from the central star and therefore require a higher excitation temperature.
Hence, we considered excitation temperatures for the rotational levels, $T_{\rm rot}$, and vibrational states, $T_{\rm vib}$, which are independent. In this way, the number of molecules in a given $v$ state and $J$ level is given by

\begin{equation}
N_{v, J} = N_{\rm AlF}(\pi r_e^2) \frac{g_J \exp(\frac{-E_J}{k_B T_{\rm rot}})} {Z_{\rm rot}(T_{\rm rot})} \:\: \frac{g_v \exp(\frac{-E_v}{k_B T_{\rm vib}})}{Z_{\rm vib}(T_{\rm vib})},
\label{NvJ}
\end{equation}
where 
$g_{\rm J}=(2J+1)$ and $g_v=1$ are the rotational and vibrational degeneracy, $E_J$ and $E_{\nu}$ are the rotational and vibrational excitation energies,
and $Z_{\rm rot}$ and $Z_{\rm vib}$ are the rotational and vibrational partition functions and are given by
%$Z(T_{\rm rot})$ is calculated using: 
\begin{equation}
Z_{\rm rot}(T_{\rm rot}) = \sum_{J}(2J+1)e^{-J(J+1)B/k_B T_{\rm rot}},
\end{equation}
where we considered that $J$ runs from 0 to 94 and B=16488.355 MHz, which is taken from CDMS \citep{Yousefi18}.

\begin{equation}
Z_{\rm vib}(T_{\rm vib}) = \sum_{v}e^{-E_v/k_B T_{vib}};
\end{equation}
here, the vibrational state $v$ runs from 0 to 5, which are the levels available in CDMS.
We varied $N_{\rm AlF}$ and $T_{\rm vib}$ in Eq. \ref{NvJ} to obtain the best fit model to all observed lines in regions 1 and 2.
We find that a vibrational temperature of $T_{\rm vib}=1300\pm500$ K with associated $Z_{\rm vib}$=1.71 and column densities of AlF molecules $N_{\rm AlF}=(5.8\pm2.0)\times10^{15}$ cm$^{-2}$ and $N_{\rm AlF}=(3.0\pm0.7)\times10^{16}$ cm$^{-2}$ reproduces the flux density of all observed line in regions 1 and 2, respectively (see dashed red lines in Figs. \ref{Mira-AlF-R1} and \ref{Mira-AlF-R2-3}). These best models are selected based on a chi-square analysis and the results are summarised in Table \ref{Mira-results} (Model C).

We note that if we include the $v=1$ data point in the calculation of the excitation temperature and column density in the population diagram shown in Fig. \ref{Mira-RD}, this results in an excitation temperature of $T_{\rm rot}=1332\pm794$ K and a column density of $N_{\rm AlF}=(2.7\pm0.7) \times 10^{16}$ cm$^{-2}$ for region 1. The derived temperature is in agreement with the vibrational temperature that we derived from separating the rotational and vibrational temperature above; however, the column density is higher by a factor of four assuming the same emitting region, which results in an overestimation of the flux density of all observed lines. For region two, it results in $T_{\rm rot}=956\pm100$ K and a column density of $N_{\rm AlF}=(7.5\pm0.4) \times 10^{16}$ cm$^{-2}$ for region two which also results to an overestimation of the flux density of observed lines (Model B in Table. \ref{Mira-results}).

The AlF lines in region 3 are also not very optically thick, with optical depths at the line centre of 0.8 and 0.02 for $v=0$ and 1 lines, respectively, calculated using Eq. \ref{opacity}.
For region 3, we assumed the same rotational and vibrational temperatures of $T_{\rm rot}=320$ K and $T_{\rm vib}=1300\pm500$ K as we found in region 2. This results in an AlF column density of $N_{\rm AlF}=(5\pm2)\times10^{16}$ cm$^{-2}$. The model results overlaid with the observed spectra are presented in the lower panel of Fig. \ref{Mira-AlF-R2-3}.
The summary of the derived column densities in all regions are listed in Table \ref{Mira-results}.

%%===========================================================================
%%===========================================================================

%%===========================================================================
%%===========================================================================

%%===========================================================================
%%=========================================================================

%===========================================================================

\begin{table*}[]
\caption{Results of excitation analysis of AlF towards $O$ Ceti in three selected regions shown in Fig. \ref{Mira-AlF-inner}.}
  \centering
  \setlength{\tabcolsep}{4.0pt}
\begin{tabular}{ccccccccc}
\hline
region & diameter  & $T_{\rm rot}$ & $T_{\rm vib}$ &  $N_{\rm AlF/H_2}$  & $N_{\rm H_2}^*$ & $f_{\rm AlF/H_2}$ & Model\\
  & ($R_\star$)     & (K) & (K) & ($\rm cm^{-2}$) & ($\rm cm^{-2}$)\\ 
\hline
1 & 22 & $145\pm40$ & $145\pm40$ & $(3.6\pm1.1)\times10^{15}$ &  & & A \\
1 & 22 & $1332\pm794$ & $1332\pm794$ & $(2.7\pm0.7)\times10^{16}$ &  && B\\
1 & 22 & $145\pm40$ & $1300\pm500$ & $(5.8\pm2.0)\times10^{15}$ & $2.4\times10^{23}$ & $(2.5\pm1.7)\times10^{-8}$ & C\\
\hline
2 & 11 & 320 & 320 & $2.1\times10^{16}$ & & & A\\
2 & 11 & $956\pm100$ & $956\pm100$ & $(7.5\pm0.4)\times10^{16}$ & & & B\\
2 &  11 & 320 & $1300\pm500$ & $(3.0\pm0.7)\times10^{16}$ & $2.0\times10^{24}$ & $(1.5\pm0.8)\times10^{-8}$ & C\\
\hline
3 & 4  & 320 & $1300\pm500$ & $(5.0\pm2.0)\times10^{16}$ & $6.1\times10^{24}$ & $(0.8\pm0.5)\times10^{-8}$ & C\\
\hline
\end{tabular}
\label{Mira-results}
\tablefoot{For regions 1 and 2: models A include $v$\,=0\ lines in the population diagram (PD) to calculate the excitation temperature and column density which  result to underestimation of the flux density of $v$\,=1\ lines as shown by blue dashed lines in Figs. \ref{Mira-AlF-R1} and \ref{Mira-AlF-R2-3}, models B include both $v$\,=0\ and $v$\,=1\ lines in the PD, which results in the overestimation of the flux densities of all lines, and model C separates $T_{\rm rot}$ and $T_{\rm vib}$ and results in the best model outcomes, as shown in Figs. \ref{Mira-AlF-R1} and \ref{Mira-AlF-R2-3} with red and green dashed lines and discussed in Sect. \ref{AlFinMira}. $^*$ We note that $N_{\rm H_2}$ is uncertain by a factor of two, as discussed in Sect. \ref{H2inMira}.} 
\end{table*}

%===========================================================================

\subsubsection{AlF/H$_2$ fractional abundance}\label{H2inMira}

Estimating the AlF/H$_2$ fractional abundance is subject to a relatively larger uncertainty because the gas density and temperature distributions in the extended atmospheres and wind-acceleration regions is variable and complex with regard to time. 

Moreover, Mira is in a binary system including Mira A ($o$ Ceti) and a white dwarf (Mira B) at a distance of $\sim0.5^{\prime\prime}$ \citep{Ramstedt14}. It is well known that the binary companion affects the gas density at large scales ($5^{\prime\prime}-10^{\prime\prime}$) \cite{Ramstedt14}. However, the gravitational field is expected to be dominated by Mira A up to $\sim 0.3^{\prime\prime} \sim 20 R_\star$ in the inner most region \cite{Mohamed12}. Therefore, the binary companion is not expected to influence the gas density in the regions we study here.

The radiation field is likely strongly affected by Mira B in the UV spectral region, even close to Mira A as suggested by the relative UV brightness of the two sources reported by \cite{Karovska97}. This probably affects the molecular abundances in the inner CSE, but how the different parameters of binary systems influence the abundances of specific molecules is not yet established. This is a topic of ongoing research \cite[e.g.][]{Saberi18, Saberi19, VandeSande22}.
We speculate that the asymmetry seen in the AlF molecular distribution ($J=10-9$) seen in Fig. \ref{Mira-AlF-inner} could be due to the gravitational influence of Mira B, but investigating this is beyond the scope of this paper.
%Mira B is the edge of region that Mira B affects the gas density . }

To constraining the H$_2$ gas column density, we used the spatially resolved C$^{18}$O($v$\,=\,0, $J$\,=\,\mbox{3--2}) line observed with ALMA. 
%To estimate the AlF abundance,
We calculated the H$_2$ column density towards the three different regions described in Section \ref{AlFinMira}.
%The H$_2$ column density was estimated using the spatially resolved C$^{18}$O ($v$\,=0\, $J$\,=\,\mbox{3--2}) line in the three regions which are marked in Fig. \ref{Mira-AlF-inner}. 
To convert the C$^{18}$O column density to H$_2$ column density, we assumed the typically used CO fractional abundance $f_{\rm C^{16}O/H_2}=4\times10^{-4}$ for M-type AGB stars \citep[e.g.][]{Khouri18}, and a ratio $\rm^{16}O/^{18}O=282 \pm 100$ reported for $o$~Ceti by \cite{Hinkle16}. 
Together, these imply a fractional abundance of C$^{18}$O, $f_{\rm C^{18}O/H_2}=(1.4\pm0.6)\times10^{-6}$.

%Together, these imply a fractional abundance of C$^{18}$O, f_{\rm C^{18}O/H_2}=(1.4^{+0.8}_{-0.4})\times10^{-6}$.}

For the excitation analysis of the C$^{18}$O line, we used the RADEX\footnote{http://var.sron.nl/radex/radex.php} radiative transfer code. We considered a line width of 6 km s$^{-1}$ and varied the excitation temperature and C$^{18}$O column density to reproduce the observed C$^{18}$O flux from observations. Calculations of the $\rm H_2$ gas density in the three selected regions are given below.

%All observed lines are extracted from a region with diameter $0.5^{\prime\prime}$ except the $J$\,=\,\mbox{4--3} line, which is extracted from a smaller region with diameter $0.25^{\prime\prime}$ to decrease the noise level. As discussed in Section~\ref{sec:Mira}, we found AlF column density $N_{\rm AlF}=(6\pm2)\times10^{15}$ cm$^{-2}$ for the best fit results as shown in Fig. \ref{Mira-AlF-R1}. 
%As discussed in Section~\ref{sec:Mira}, using the population diagram method we found $T_{\rm rot}=145\pm 40$ K, $T_{\rm vib}=1300\pm500$ K, and $N_{\rm AlF}=(6\pm2)\times10^{15}$ cm$^{-2}$ for the best fit results for all the observed lines. The observed lines overlaid with the results are presented in Fig. \ref{Mira-AlF-R1}. 
In region~1, the brightness temperature of the C$^{18}$O line peaks at 13 K, corresponding to a flux density of 90 mJy as shown in Fig. \ref{Mira-CO}.
We varied the C$^{18}$O excitation temperature and column density for C$^{18}$O to reproduce the observed flux density. Assuming an excitation temperature of $700\pm300$ K, we found that a column density $\rm N_{C^{18}O}=(3.3\pm1.3)\times10^{17}$ cm$^{-2}$ reproduces the observed line. %for the best model based on the chi-2 calculations. 
%These gives H$_2$ column densities $N_{\rm H_2}=(2.4\pm0.8)\times10^{23}$ cm$^{-2}$. This implies an AlF fractional abundance within $f_{\rm AlF/H_2}\sim(2.5\pm1.6)\times10^{-8}$ in region 1.
Considering $f_{\rm C^{18}O/H_2}=(1.4\pm0.6)\times10^{-6}$ gives the H$_2$ column densities $N_{\rm H_2}=2.4\times10^{23}$ cm$^{-2}$. 
 The derived $N_{\rm H_2}$ has an uncertainty of a factor of 2.1 considering the uncertainty on the $\rm^{16}O/^{18}O$ isotopic ratio and excitation temperature.
Assuming the AlF column density of $N_{\rm AlF}=(5.8\pm2)\times10^{15}$ cm$^{-2}$ that we describe in the previous section, we find an AlF fractional abundance of $f_{\rm AlF/H_2}\sim(2.5\pm1.7)\times10^{-8}$ in region 1.

% Region 2: 
%we use the spatially resolved $v$\,=0\,, $J$\,=\,\mbox{4--3} and $J$\,=\,\mbox{10--9}, and $v$\,=1\,, $J$\,=\,\mbox{7--6} lines extracted from the region with the diameter of 0.168$^{\prime\prime}$ shown in Fig. \ref{Mira-AlF-inner}. We find that a column density $N_{\rm AlF}=(3.0\pm0.7)\times10^{16}$ cm$^{-2}$ will reproduce all the observed lines as shown in Fig. \ref{Mira-AlF-R2}. 
In region~2, the brightness temperature of the C$^{18}$O line peaks at 30 K, corresponding to a flux density of 55 mJy.
Similarly to region 1, we varied the excitation temperature and the column density of C$^{18}$O to reproduce the line flux extracted from region 2.
We assumed an excitation temperature of $1500\pm500$ K and found a column density of $\rm N_{C^{18}O}=(1.5\pm0.5)\times10^{18}$ cm$^{-2}$ that reproduces the C$^{18}$O line.
These yield a H$_2$ column density of $N_{\rm H_2}=1.1\times10^{24}$ cm$^{-2}$ with an uncertainty of a factor of 1.8 due to the uncertainties on the isotopic ratio and the excitation temperature. 
For the same region, \cite{Khouri18} reported $N_{\rm H_2} =3.4\times10^{24}$~cm$^{-2}$ based on the radiative transfer modelling of the CO ($v$\,=\,1, $J$\,=\,\mbox{3--2}) and $^{13}$CO~($v$\,=\,0, $J$\,=\,\mbox{3--2}) line.
Thus, a column density of $N_{\rm H_2} \sim 2\times10^{24}$ would be consistent with both the value determined by us and those by \cite{Khouri18} given the intrinsic uncertainties and different approaches, and is probably a better estimate of the real column density.
The best model from Sect. \ref{AlFinMira} gave an AlF column density $N_{\rm AlF}=(3.0\pm0.7)\times10^{16}$ cm$^{-2}$.
Therefore, the estimated H$_2$ and AlF column densities result to a fractional abundance of $f_{\rm AlF/H_2}\sim(1.5\pm0.8)\times10^{-8}$ in region 2.

In region~3, the C$^{18}$O brightness temperature peaks at 160 K corresponding to a flux density of 38 mJy.
We assumed an excitation temperature of $1500\pm500$ K and found a column density of $(8.5\pm2.5)\times10^{18}$ cm$^{-2}$ for C$^{18}$O to reproduce the observed brightness temperature in this region.
This implies a H$_2$ column density of $N_{\rm H_2}=6.1\times10^{24}$ cm$^{-2}$ with an uncertainty of a factor of 2.0. From Sect. \ref{AlFinMira}, we find an AlF column density of $\rm N_{AlF}=(5\pm2)\times10^{16}$ cm$^{-2}$ for the best model.
These translate to a fractional abundance of $f_{\rm AlF/H_2}\sim(0.8\pm0.5)\times10^{-8,}$ and in region 3 this is within $4R_{\star}$. The derived fractional abundances in the three regions are listed in Table \ref{Mira-results} (Model C).

%We find the H$_2$ column density to be $(2.4\pm1.4)\times10^{23}$ cm$^{-2}$, $(1.1\pm0.6)\times10^{24}$ cm$^{-2}$, and $(6.1\pm3.2)\times10^{24}$ cm$^{-2}$ in regions 1, 2, and 3, respectively. These translate to AlF fractional abundances of $f_{\rm AlF}\sim(2.5\pm1.7)\times10^{-8}$, $(2.7\pm1.6)\times10^{-8}$, and $(0.8\pm0.5)\times10^{-8}$ for regions 1, 2, and 3.

Our derived AlF fractional abundances in the three regions are consistent with chemical models of \citet[][ see Fig. A.4,]{Agundez20} which also predict a mean AlF fractional abundance of $\sim10^{-8}$ within $\sim9R_\star$ and also a lower abundance in a range of $\sim10^{-11}-10^{-8}$ in the innermost region with a radius of $1R_\star<R<3R_\star$. Considering the initial mass of $o$ Ceti $\sim1 M_\odot$, our results are also in agreement with the stellar yield models by \cite{Lugaro04, Karakas10} and are consistent with the Solar F budget of F/H$_2=(5\pm2)\times10^{-8}$ \citep{Asplund21}.

\subsubsection{The excitation of AlF}
\label{sec:excitation}

The derived rotational temperatures of 320~K and 145~K from the population diagram in the inner CSE are rather low and seem to indicate sub-thermal excitation of AlF. A detailed study of the excitation of AlF is necessary to understand the distribution of the level populations. However, such an analysis is complicated by the fact that the radiation field as a function of position is poorly constrained at the relevant wavelengths (because of, for example, dust absorption and emission). 
We do not expect the binary companion to have a significant effect on the radiation field close to Mira~A at the relevant wavelengths. 

Although, the three-dimensional gas density distribution in the inner region of $o$~Ceti is only constrained by one-dimensional models. Nonetheless, we were able to estimate the relative effects of collisions and the radiation field on the excitation of AlF. 

The radiative pumping of AlF from $v=0$ to $v=1$ takes place through an infrared band at 12.48~$\mu$m. Assuming 
the effective temperature and effective near-infrared (IR) radius as determined by \cite{Wittkowski2016}, we can estimate the 12.48~$\mu$m flux density for a 2450~K black body, and we find a flux density of $\sim 500$~Jy. This does not include the contribution from diffuse emission from dust and the binary companion and, hence, is a lower limit to the actual radiation field.
For comparison, this naked-star estimate corresponds to 0.23 times the flux taken directly from the infrared space observatory (ISO) observations\footnote{https://irsa.ipac.caltech.edu/data/SWS/}. 

To compute the mean intensity at 12.48~$\mu$m from a naked star with the considered radius and effective temperature, we calculated the dilution of the radiation field over the solid angle of region~1 ($\Omega=2.134\times 10^{-12}$~sr). We find a mean intensity of $J_\nu=F_\nu/\Omega=1.04\times 10^{-8} \rm ergs \:s^{-1}$ cm$^{-2}$ Hz$^{-1}$ sr$^{-1}$. The IR pumping rate can be written as $\rho = A_{\rm vib} J_\nu / (2hc(\nu/c)^3) = 0.1 {\rm s}^{-1}$, where $A_{\rm vib} \sim 8.62$ s$^{-1}$ is the spontaneous-emission coefficient of AlF for a $v=1-0$ transition. Considering collisional rates of $\sim 2 \times 10^{-10}$~cm$^3$ s$^{-1}$ \citep{Danilovich21}, we find a upper limit for the H$_2$ densities below which IR pumping dominates by 10$^9$~cm$^{-3}$. If the radiation field is stronger than our naked-star estimate, the IR pumping will be efficient at larger densities. Given the observed ISO spectrum integrated over a much larger area than region 1, the mean intensity is likely at most a factor of a few larger than our estimate.

Assuming depths for regions~1 and 2 along the line of sight similar to their radial extent on the sky ($\sim 5 \times 10^{14}$~cm and $\sim 3 \times 10^{14}$~cm, respectively), the H$_2$ column densities we obtained imply average densities in these regions of $\sim 5\times 10^8$~cm$^{-3}$ and $\sim 4 \times 10^{9}$~cm$^{-3}$.
Hence, the radiation field is expected to dominate the excitation through vibrational pumping in a large fraction of region~1 but not as much in region~2. Interestingly, IR pumping would be expected to help increase the rotational temperature of the $v=0$ levels for low gas densities, making the low values we derive puzzling. In order to study the excitation of AlF in detail, radiative transfer models including the three-dimensional 
gas-density distribution and mean intensity at relevant frequencies is necessary. Nonetheless, the rotational and vibrational temperatures we derive provide an empirical description of the average excitation of AlF and can be used to infer column densities as above.

\subsubsection{HF}

As discussed in Sect. \ref{Introduction}, HF is among the two most abundant F-bearing species in the outflow of AGB stars \citep{Agundez20}.
We investigated the archive HF data available for $o$ Ceti, aiming to determine the HF fractional abundance and the total F budget in the gas phase. There is a tentative detection of HF ($J$\,=\,\mbox{1--0}) at 1232.476 GHz observed with Herschel/SPIRE. There is also potential detection of the HF lines ($J$\,=\,\mbox{2--1}, \mbox{3--2}, and \mbox{4--3}) observed by PACS in Herschel. However, all these lines are blended with H$_2$O lines, making it difficult to determine the HF fractional abundance based on low spectral resolution of Herschel data. % for M-type AGB stars which have high abundance of H$_2$O molecule.

%===========================================================================
\subsection{R Leo}\label{RLeo}
%===========================================================================

In R Leo, the AlF ($v$\,=\,0, $J$\,=\,\mbox{7--6}) line is detected at five different epochs with ALMA at various pulsation phases. The data are taken over a period of one year at a visual phase of $\phi=$ 0.0, 0.5, 0.6, 0.7, and 0.9.
One extra line of AlF ($v$\,=\,0, $J$\,=\,\mbox{15--14}) is covered by the ACA observations.
All the covered lines with their spectroscopic parameters and measured intensities are listed in Table \ref{AlF}.
The observations show that the emission region is barely resolved by the $\sim0.13^{\prime\prime}$ beam (Fig. \ref{RLeo-image}). The line flux extracted from the higher angular-resolution images (beam $\sim0.02^{\prime\prime}$) increases up to apertures of $\sim0.25^{\prime\prime}$ . Hence, we considered the emission region to have a diameter of $\sim 0.25^{\prime\prime}\sim 33$ au $\sim 18R_{\star}$.
%We constrained the emission region by extracting the line spectrum from increasingly larger regions until the line flux stopped increasing. Based on this, we find and emission region with a diameter of ~0.3'', ~ 40 au, ~R_\star.

\begin{figure*}[tb]
\raggedright
\includegraphics[width=160mm]{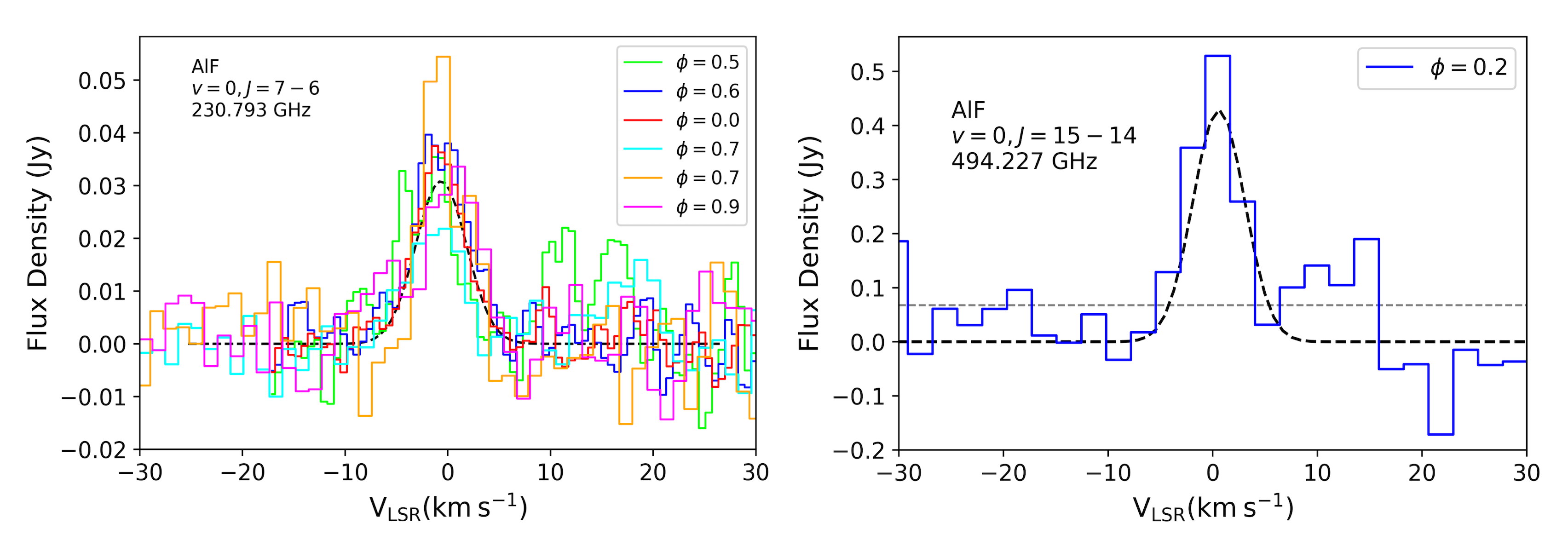}
 \caption[]{\label{}
    ALMA observations of AlF line emission towards R Leo (solid lines) and the results of an LTE model (dashed black lines). Stellar variability phases ($\phi$) are indicated in the upper-right, and the line transitions are indicated in the upper-left panels. The grey dashed lines show the rms level.}
\label{RLeo-multi-AlF}  
\end{figure*}

Figure \ref{RLeo-multi-AlF} presents the multi-epoch observations of AlF at 230.79 GHz. 
Our observations with the highest angular resolution at $\phi=$0.7 and 0.9 are subject to flux loss due most likely to the limited maximum recoverable scale of $0.4-0.6^{\prime\prime}$ shown in Fig. \ref{RLeo-multi-AlF}, left panel, with magenta and cyan profiles. The flux variation seen in the AlF ($J$\,=\,\mbox{7--6}) line is most likely due to low surface brightness sensitivity of the long baseline observations and the added limitations related to imaging emission with an extent similar to the maximum recoverable scale. Possible small calibration uncertainties on the shortest baselines and changes in antenna configuration between the observations makes a direct comparison between the highest angular resolution observations uncertain.
Therefore, based on the current observations, we cannot confirm any flux variation due to the stellar variability.

\begin{figure}[tb]
 \centering
\includegraphics[width=80mm]{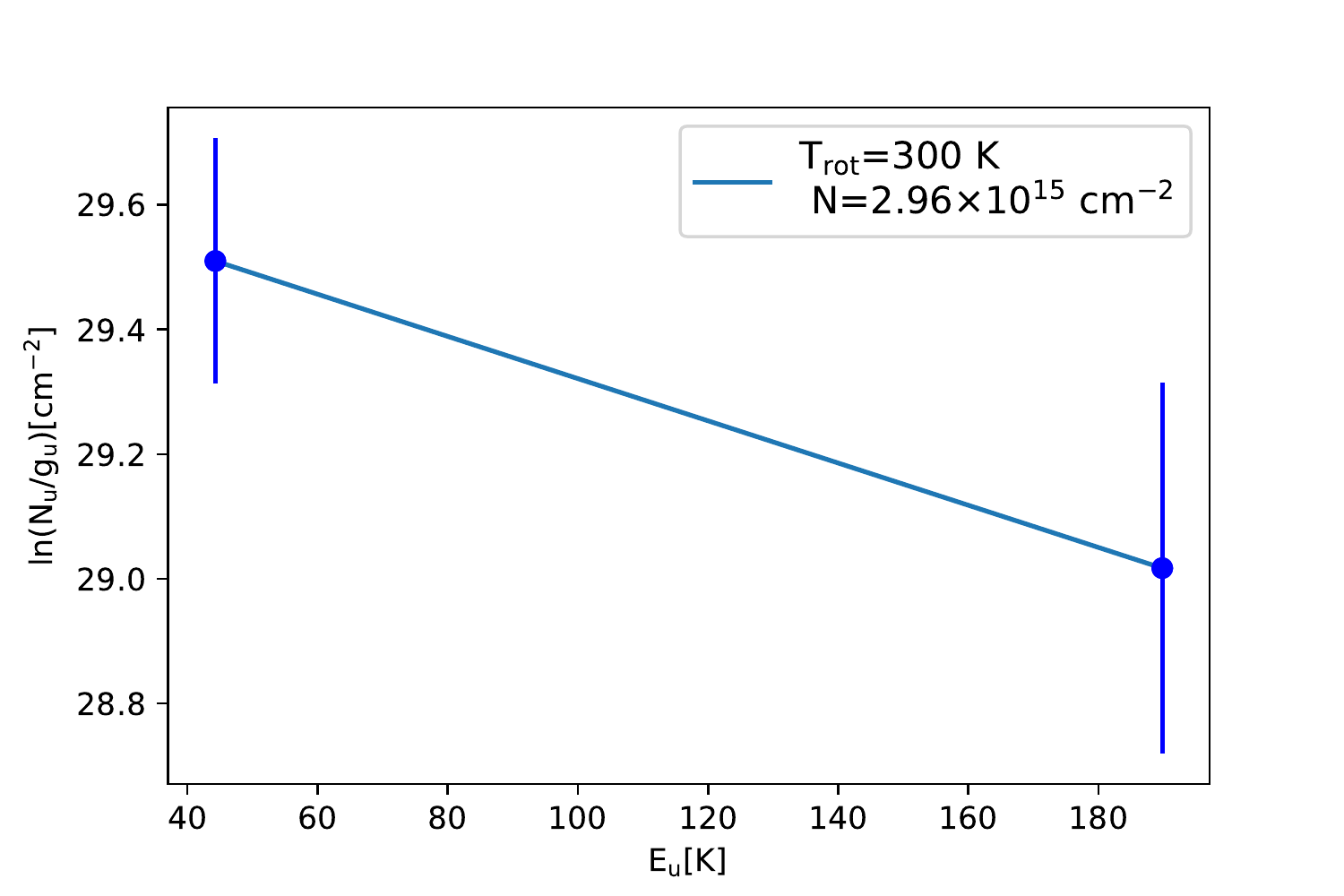}
 \caption[]{\label{}
     Population diagram of observed AlF rotational lines towards R Leo. The data point with $E_u=44$ K is the mean of the first three observations that are listed in Table \ref{AlF} (see Sect. \ref{RLeo} for explanations).}
\label{RLeo-RD}  
\end{figure}

For the population diagram, we used the weighted mean value of the first three $J$\,=\,\mbox{7--6} data points that are listed in Table \ref{AlF}. We removed the three observations with MRS $<0.6^{\prime\prime}$ that are likely subject to flux loss.
As shown in Fig. \ref{RLeo-RD}, we derived the rotational temperature of \mbox{$T_{\rm rot}\sim300$\,K} and the column density of $N_{\rm AlF}\sim3.0\times10^{15}$ cm$^{-2}$. %$N_{\rm AlF}\sim(2.0\pm0.5)\times10^{15}$ cm$^{-2}$. 
The accuracy of the results is limited since we only have two points in the population diagram.
We used the column density and the rotational temperature derived from the population diagram to reproduce the observations. The observed spectra and the model results are presented in Fig. \ref{RLeo-multi-AlF}. 

Since R Leo and $o$ Ceti have similar mass-loss rates and pulsation periods, we assumed the same H$_2$ column density of $(2.4\pm1.4)\times10^{23}$ cm$^{-2}$ as we derived for a region with a radius of 17 au in $o$ Ceti (region 1). We also crudely estimated an AlF fractional abundance of $f_{\rm AlF/H_2}\sim (1.2\pm0.5)\times 10^{-8}$ for R Leo. 
%A more precise determination of the AlF fractional abundance requires 

%This is a crude approximation due to the large uncertainty of the H$_2$ column density in the emitting region.

%%===========================================================================
%%===========================================================================
%======================================================

%%===========================================================================
%%===========================================================================

\subsection{R Dor, W Hya, and IK Tau}\label{3sources}
%======================================================

We report a tentative detection of AlF ($v$\,=\,0, $J$\,=\,\mbox{15--14}) line emission at 494.223 GHz towards W Hya and R Dor. 
We also report tentative detection of AlF ($v$\,=\,0, $J$\,=\,\mbox{7--6}) line emission at 230.79 GHz towards IK Tau and W Hya.
The observed lines with their properties extracted from Gaussian fits are listed in Table \ref{AlF}.
We assumed an emitting region of $11R_{\star}$, where $R_\star$ refers to the stellar radius measured in infrared as listed in Table. \ref{Source}, for these sources similar to what we derived from observations of $o$ Ceti. This results in an emitting region with a radius of 22 au for W Hya, 13.5 au for R Dor, and 17.5 au for IK Tau.

We considered $T_{\rm rot}$ of 145 K from Mira region one and 300 K from Mira region two and R Leo to estimate the mean column density in these sources. The results are listed in Table. \ref{AlFinOthersources}.
%Assuming a rotational temperature of 300 K, as found for R Leo, we derived the AlF column densities to be $(1.5\pm0.5)\times10^{15}$ cm$^{-2}$ for W Hya, $(3.0\pm0.5)\times10^{14}$ cm$^{-2}$ for R Dor, and $(1.5\pm0.5)\times10^{16}$ cm$^{-2}$ for IK Tau. The model results overlaid on the observed spectra are presented in Fig. \ref{AlF-3S}.
The model results overlaid on the observed spectra are presented in Fig. \ref{AlF-3S}.

To estimate the fractional abundances for these sources, we need to know the H$_2$ column density. Since AlF emission is compact and it comes from a region that can still be gravitationally bound to the star, part of the gas will be ejected and some will fall back to the star. Hence, estimating gas densities in these inner regions from extrapolating models obtained based on the large-scale envelope will lead to errors in the estimated gas densities. Moreover, densities from models for the large-scale envelopes are uncertain by a factor of a few \cite{Ramstedt08}. Finally, the way in which the densities in the extended atmosphere differ from those predicted by extrapolating the mass-loss rate inwards may vary between stars because of differences in the extended atmospheres and wind-acceleration region \cite[e.g.][]{Habing96, Hofner18}. 
%Therefore to approximate the order of magnitude of the AlF fractional abundance, we used the H$_2$ column density that we derived for $o$ Ceti.

To roughly estimate the AlF fractional abundances in these sources, we used the H$_2$ column density that we derived for $o$ Ceti of $(2.4\pm1.4)\times10^{23}$ cm$^{-2}$ within a radius of $11R_\star$ from the central star.
This suggests the $\rm AlF/H_2$ fractional abundance is within a range of $\sim(0.1-4)\times10^{-8}$ for these sources as listed in Table \ref{AlFinOthersources}. This is in agreement with the Solar fractional abundance of F and the stellar yield models for AGB stars with initial masses in range of $1-2 M_\odot$.
%we can make roughly estimate the AlF fractional abundance of $\sim(0.6\pm0.4)\times 10^{-8}$ for W Hya, $\sim(1.3\pm0.8)\times 10^{-9}$ for R Dor, and $\sim(6.2\pm4.2)\times 10^{-8}$ for IK Tau.
We note that the approximation we made above is crude. This is even less certain for IK Tau and R Dor due to different mass-loss rates (in case of IK Tau) and pulsation periods (in both cases) with respect to $o$ Ceti, which may cause even larger differences in the H$_2$ gas density and fractional abundances in the inner region. A proper determination of AlF abundances will require a study of gas densities in the AlF emission regions.

%===========================================================================
\begin{table}[]
\caption{Approximation of AlF column density and fractional abundance in other sources.}
  \centering
  \setlength{\tabcolsep}{0.5pt}
\begin{tabular}{c|cc|cc} %|l|l|l|l|
\hline
 Star& \multicolumn{2}{c|}{$T_{\rm rot}=145$ (K)} & \multicolumn{2}{c}{$T_{\rm rot}=300$ (K)} \\
% \hline
 &  $N_{\rm AlF}$ ($\rm cm^{-2}$)  & $f_{\rm AlF/H_2}$  & $N_{\rm AlF}$ ($\rm cm^{-2}$)  & $f_{\rm AlF/H_2}$ \\
\hline
R Leo &  &  & $3\times10^{15}$ & $1.2\times10^{-8}$\\
R Dor &  $2.5\times10^{14}$ & $0.1\times10^{-8}$ & $3\times10^{14}$ & $0.12\times10^{-8}$\\
W Hya &  $1.2\times10^{15}$ & $0.5\times10^{-8}$ & $1.5\times10^{15}$ & $0.6\times10^{-8}$\\
IK Tau &  $0.6\times10^{16}$ & $2.5\times10^{-8}$ & $1\times10^{16}$ & $4.2\times10^{-8}$\\
\hline
\end{tabular}
\label{AlFinOthersources}
\tablefoot{For R Leo, the $T_{\rm rot}$ and $N_{\rm AlF}$ are estimated using the PD shown in fig. \ref{RLeo-RD}. For other sources, we used $T_{\rm rot}$ estimated for $o$ Ceti and R Leo to approximate $N_{\rm AlF}$. For all sources, the H$_2$ density of $2.4\times10^{23} \rm cm^{-2}$ is used as derived for $o$ Ceti.} 
\end{table}
%===========================================================================

%{\bf This estimate is less certain for R Dor and IK Tau due to the differences in pulsation periods with $o$ Ceti.}
%===========================================================================
\section{Discussion and summary}
%===========================================================================

The cosmic origin of F is still uncertain. AGB stars are among the few candidates to synthesis F in our Galaxy. %, however their contribution is not clear. 
From stellar yield models, the efficiency of F synthesis in AGB stars strongly depends on the initial mass and metallicity \citep{Lugaro04, Karakas10}. For Solar metallicity, the F synthesis is maximal for stars with an initial mass of $2-4 M_\odot$. From chemical models by \cite{Agundez20}, a significant amount of F is expected to be locked into AlF and HF in the outflow of all chemical types of AGB stars.
In this paper, we report the first detection of AlF line emission towards five oxygen-rich AGB stars observed
with ALMA: $o$ Ceti, R Leo, W Hya, R Dor, and  IK  Tau.

Towards $o$ Ceti, we detected five rotational lines and determined the fractional abundance of $f_{\rm AlF/H_2}\sim(0.8\pm0.5)\times10^{-8}$ within a radius of $2R_\star$, $\sim(1.5\pm0.8)\times10^{-8}$ within a radius  of$5.5R_\star$ ,and $\sim(2.5\pm1.7)\times10^{-8}$ within a radius of $11R_\star$ from population diagram analysis. The observations are best reproduced by considering independent rotational and vibrational excitation temperatures.
These derived fractional abundances at various radii from the star are in agreement with $f_{\rm AlF/H_2}$ molecular distribution for an M-type AGB star from the recent chemical models by \cite{Agundez20}. This indicates how spatially resolved observations of several transitions can verify the accuracy of chemical models on predictions of molecular fractional abundances as long as the H$_2$ gas density in the emitting region is known.

Towards R~Leo, we find a column density of $3\times10^{15}$ cm$^{-2}$ for an emission region with radius $\sim 9 R_\star$. 
For other sources, we considered the rotational temperatures of 145 K and 300 K as derived for $o$ Ceti and R Leo to make a rough estimation of the AlF column density. These result in $N_{\rm AlF}\sim(1.2-1.5)\times10^{15}$~cm$^{-2}$ in W Hya, $\sim(2.5-3.0)\times10^{14}$~cm$^{-2}$ in R Dor, and $\sim(0.6-1.0)\times10^{16}$~cm$^{-2}$ in IK Tau within a radius of $11R_{\star}$ from central stars.
However, spatially resolved observations towards these sources are necessary to resolve the line emitting regions and constrain the column densities. 
By assuming the same H$_2$ column density as we derived for $o$ Ceti, we can make a crude approximation of the AlF fractional abundance $(1.2\pm0.5)\times10^{-8}$ in R Leo and in a range of $(0.1-4)\times10^{-8}$ for W Hya, R Dor, and IK Tau.

All observed sources in our sample have an initial mass in the range of $M\sim1-2M_{\odot}$ and are Galactic sources with metallicities probably similar to solar; thus, they are not expected to efficiently synthesise fluorine from stellar yield models by \cite{Lugaro04, Karakas10}.
Our results for all sources are in a good agreement with both stellar yield models and chemical models.

\cite{Danilovich21} recently reported the detection of AlF and HF towards the S-type AGB star, W Aql. Using radiative transfer analysis, they found fractional abundances of $ f_{\rm AlF/H_2}=1\times10^{-7}$ and $ f_{\rm HF/H_2}=1\times10^{-8}$. Their reported value in the inner part is a higher than expected AlF abundance for W Aql, which has an initial mass within a range of $1.2-1.6 M_{\odot}$ reported by \cite{DeNutte17}. 
This can indicate that either the mass of W~Aql is larger, or that models for fluorine production predict nucleosynthesis at initial masses that are too large. However, the uncertainty in the radiative transfer analysis based on a single line observation and uncertainties in the physical parameters in the inner CSEs can play an important role in the molecular excitation analysis and abundance derivation.
%Their model for AlF abundance is also could not reproduce the $v=1, J=7-6$ line indicating that the $v=0$ and $v=1$ lines have likely different excitation mechanism. 

The estimated $f_{\rm AlF/H_2}$ in all M-type AGB stars in our sample are in agreement with the reported $f_{\rm AlF/H_2}$ in the Sun and the C-type AGB star, IRC+10216, reported by \cite{Asplund21} and \cite{Agundez12}, respectively.
We note that dependency of the AlF abundance on the AGB chemical type is not observationally constrained. This is due to very few observations that have been done so far to measure the abundances of F-bearing species in the outflow of evolved stars due to the lack of sensitivity of previous generations of observational facilities. Chemical models by \cite{Agundez20} have assumed the same photospheric abundance of $f_{\rm AlF/H_2}\sim 10^{-8}$ for M-, S-, and C-type AGB stars. Further observations in a larger sample from all chemical types are still needed to verify this assumption and also quantify the total F budget in various chemical types.

%Our estimated AlF fractional abundances are consistent with the solar F fractional abundance for all sources.
From chemical models, a significant overabundance of F due to stellar nucleosynthesis is expected to be seen in both AlF and HF abundances (M. Ag{\'u}ndez, priv. comm.). 
We remind the reader that the low spectral resolution of Herschel PACS/SPIRE data of HF observations make it impossible to distinguish the HF lines from the H$_2$O lines that are abundant in M-type AGB stars. 
Therefore, our study suggests that observations of AlF lines towards AGB stars with initial masses of $\mbox{2--4} M_{\odot}$ can provide more reliable observational evidence of the F nucleosynthesis predicted by stellar yield models of AGB stars. This is important with regard to understanding the role of AGB stars in the total F production in our Galaxy.

%%%%%%%%%%%%%%%%%%%%%%%%%%
\begin{acknowledgements}
%%%%%%%%%%%%%%%%%%%%%%%%%%

We thank John Black for fruitful discussions on the excitation of the AlF and all his comments on the manuscript.
This paper makes use of the following ALMA data: ADS/JAO.ALMA$\#$2016.1.00004.S, $\#$2017.1.00191.S, $\#$2018.1.00749.S, $\#$2018.1.00649.S, $\#$2016.1.01202.S, $\#$2017.1.00862.S, $\#$2016.1.00119.S, $\#$2018.1.01440.S, and $\#$2016.2.00025.S. ALMA is a partnership of ESO (representing its member states), NSF (USA) and NINS (Japan), together with NRC (Canada), MOST and ASIAA (Taiwan), and KASI (Republic of Korea), in cooperation with the Republic of Chile. The Joint ALMA Observatory is operated by ESO, AUI/NRAO and NAOJ.
MS and SW acknowledge support by the SolarALMA project, which has received funding from the European Research Council (ERC) under the European Union’s Horizon 2020 research and innovation programme (Grant agreement No. 682462), and by the Research Council of Norway through its Centres of Excellence scheme, project number 262622. TK, LVP and WV acknowledge support from the Swedish Research Council under grants No. 2019-03777 and 2014-05713. LVP acknowledges the ERC consolidator grant 614264. JPF has received funding support from the European Research Council under the European Union’s Seventh Framework Program (FP/2007-2013) / ERC Grant Agreement n. 610256 NANOCOSMOS.

%This paper makes use of the following ALMA data: ADS/JAO.ALMA$\#$2018.1.00749.S and ADS/JAO.ALMA$\#$2018.1.00649.S. ALMA is a partnership of ESO (representing its member states), NSF (USA) and NINS (Japan), together with NRC (Canada), MOST and ASIAA (Taiwan), and KASI (Republic of Korea), in cooperation with the Republic of Chile. The Joint ALMA Observatory is operated by ESO, AUI/NRAO and NAOJ. 
%In addition, publications from NA authors must include the standard NRAO acknowledgement: The National Radio Astronomy Observatory is a facility of the National Science Foundation operated under cooperative agreement by Associated Universities, Inc.

\end{acknowledgements}

%%%%%%%%%%%%%%%%%%%%%%%%%%%%%%%%
%% references

%\documentclass{aa}

%\bibpunct{(}{)}{;}{a}{}{,} % to follow the A&A style

% for the bibliography, at the end
\bibliographystyle{aa} % style aa.bst
\bibliography{refrences} % your references Yourfile.bib
%\documentclass[bibyear]{aa}

%\end{document}

%%%%%%%%%%%%%%%%%%%%%%%%%%%%%%%%%
 \begin{appendix}
% %%%%%%%%%%%%%%%%%%%%%%%%%%%%%%%%%

%===========================================================================
\section{CO observation towards $o$ Ceti}\label{COinMira}
%===========================================================================

\begin{figure}[tb]
\raggedright
\includegraphics[width=80mm]{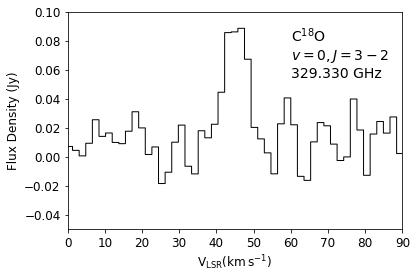}
 \caption[]{\label{}
    ALMA observations of $\rm C^{18}O$ line towards $o$ Ceti from region one. This line has been used for the estimation of H$_2$ density in region one as discussed in Sect. \ref{H2inMira}.}
\label{Mira-CO}  
\end{figure}

Figure \ref{Mira-CO} presents $\rm C^{18}O$ ($v=0, J=3-2$) line spectra extracted in region one towards $o$ Ceti. The line has been used to estimate the H$_2$ gas density in region 1 as detailed in Sect. \ref{H2inMira}.
%The gas density estimated from CO observations can have uncertainties by a factor of three which will accordingly affects the molecular fractional abundances \citep[e.g.][]{Ramstedt08}.

%%===========================================================================
%%===========================================================================

%===========================================================================
\section{R Leo}\label{App-1}
%===========================================================================

\begin{figure}[tb]
\raggedright
\includegraphics[width=80mm]{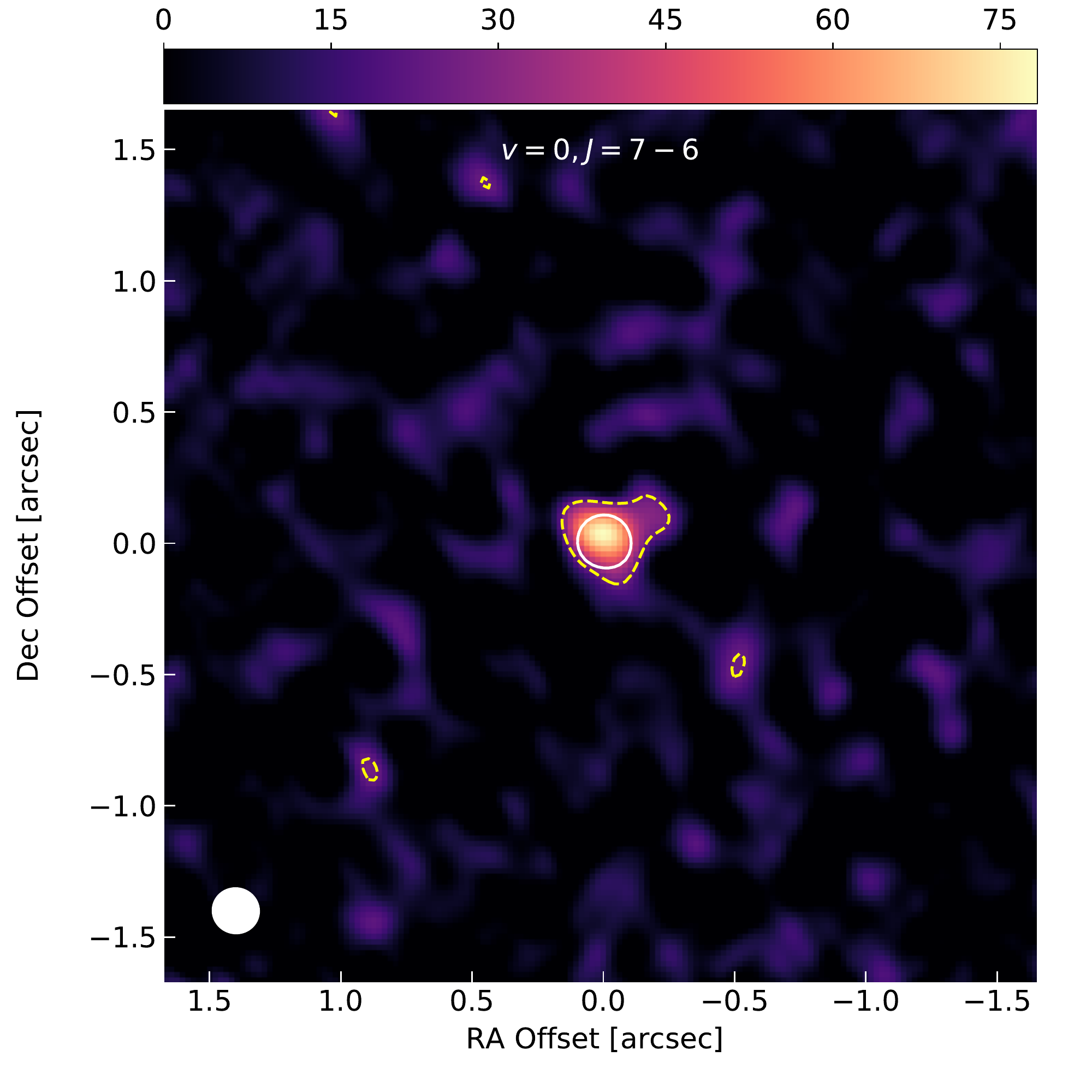}
 \caption[]{\label{}
    Integrated emission of AlF ($J=7-6$) line towards R Leo. The lines are integrated in a range of  -7 to 5 km/s. The scale is given in Jy~km/s/beam. The white contours show the 50\% stellar continuum emission level at the corresponding frequencies, and the dashed yellow contours mark the 3-$\sigma$ level of the line emission. The filled white ellipses indicate the beam size ($0.133^{\prime\prime}$) in each observation.}
\label{RLeo-image}  
\end{figure}

Figure \ref{RLeo-image} presents the integrated emission of the AlF ($J=7-6$) line observed with a $0.133^{\prime\prime}$ beam towards R Leo that is discussed in Section \ref{RLeo} as an estimator of the size of the emitting region.

%%===========================================================================
%%===========================================================================

%===========================================================================
\section{R Dor, W Hya, and IK Tau}\label{App-2}
%===========================================================================
%Figure \ref{RLeo-multi-AlF} presents the AlF observations and results of the LTE model for R Leo that are detailed in Section \ref{RLeo}.

\begin{figure*}[]
 \centering
\includegraphics[width=140mm]{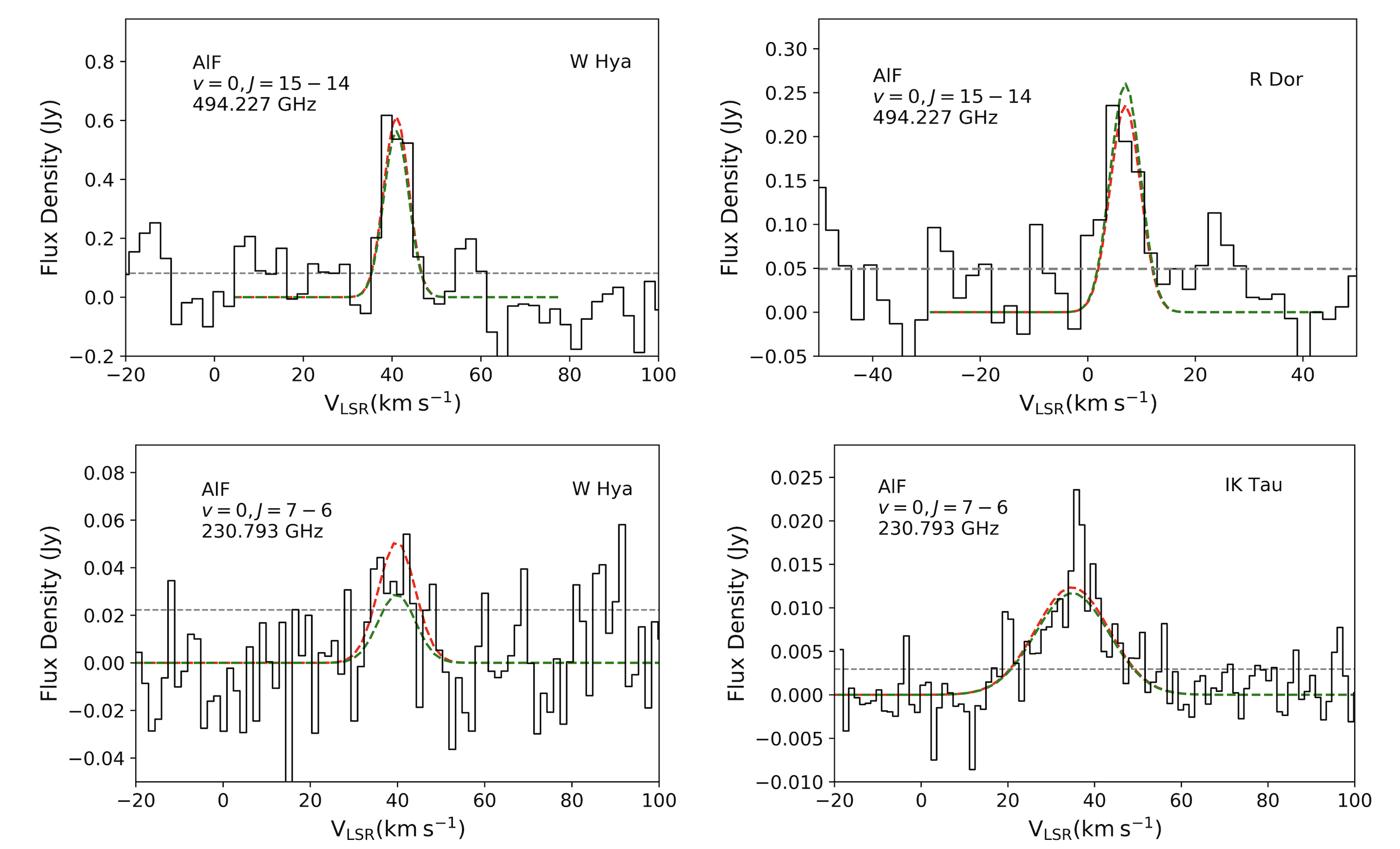} %{AlF-Spectra-3S.png}
 \caption[]{\label{}
    ALMA observations of AlF lines towards IK Tau, R Dor, and W Hya (Black solid lines) overlaid with LTE model results using $T_{\rm rot}=145$ K (red dashed lines) and $T_{\rm rot}=300$ K (green dashed lines) that are detailed in Sect. \ref{3sources}. The line rest frequencies and transition are stated in each panel. The AlF ($v$\,=\,0, $J$\,=\,\mbox{7--6}) line emission at 230.7938 GHz can potentially be blended with $\rm ^{50}TiO_2$ line at 230.7931 GHz. The grey dashed lines show the rms level.}
\label{AlF-3S}  
\end{figure*}

Figure \ref{AlF-3S} presents the observed spectra for R Dor, W Hya, and IK Tau overlaid with the model results from Section \ref{3sources}.

%%===========================================================================
%%===========================================================================

%===========================================================================
% \section{Non-detected sources}\label{App-ND}
% %===========================================================================

% %===========================================================================
% \begin{table}[]
% \caption{Upper limits of AlF lines in non-detected AGB stars with ALMA-ACA Band 8.}
%   \centering
%   \setlength{\tabcolsep}{2.5pt}
% \begin{tabular}{ccccccccc}
% \hline
% Star &  $\dot M$ & d  & $M_\star$ & $T_{\rm peak}$\\
%      & ($M_{\odot} yr^{-1}$) & (pc) & ($M_{\odot}$) & Jy
%      \\ 
% \hline
% TX psc &  $2.4\times10^{-8}$ &  275 & \\
% TW hor &  $2.4\times10^{-8}$ & 322 & \\
% U Hya & $1\times10^{-7}$ & 160 \\
% EP Aqr &  $3\times10^{-7}$ & 114 \\
% R Hya & $1\times10^{-7}$ &  124 \\
% \hline
% \end{tabular}
% \label{NonD-Source}
% \tablefoot{.} 
% \end{table}
% %===========================================================================

\end{appendix}

\end{document}